\documentclass{ws-ijmpd}
\usepackage[super,compress]{cite}
\usepackage{url}
\usepackage[dvips,breaklinks]{hyperref}
\usepackage{breakurl}

\begin{document}
\title{Constraints on thawing scalar field models from fundamental constants}
\author{Qing Gao$^1$\footnote{Email: gaoqing01good@163.com} ~ and
Yungui Gong$^{1,2}$\footnote{Email: yggong@mail.hust.edu.cn}}

\address{$^1$MOE Key Laboratory of Fundamental Quantities Measurement, School of Physics, Huazhong
University of Science and Technology, Wuhan 430074, China\\
$^2$Institute of Theoretical Physics, Chinese Academy of Sciences, Beijing 100190, China}

\maketitle

\begin{abstract}
We consider a dark energy model
with a relation between the equation of state parameter $w$ and
the energy density parameter $\Omega_\phi$ derived from thawing scalar field models. Assuming the variation of
the fine structure constant is caused by dark energy, we use the observational
data of the variation of the fine structure constant to constrain
the current value of $w_0$ and $\Omega_{\phi 0}$ for the dark energy model.
At the $1\sigma$ level, the observational data excluded some areas around $w_0=-1$,
which explains the positive detection of the variation of the fine structure constant at the $1\sigma$ level,
but $\Lambda$CDM model is consistent with the data at the $2\sigma$ level.
\end{abstract}



\section{Introduction}

The observations of type Ia supernovae in 1998 \cite{acc1,acc2} found that the universe is experiencing an accelerating expansion.
To explain the cosmic acceleration, an exotic energy with negative pressure which contributes approximately 70\% of the energy density in the universe was introduced to the matter component.
This exotic energy component is called dark energy. For a recent review of dark energy, please see Refs. \refcite{EJ,limiaode}.
We usually use a minimally coupled scalar field with positive kinetic energy term to model the quintessence \cite{wetterich88,peebles88,quintessence},
and a minimally coupled scalar field with negative kinetic energy term to model the phantom \cite{phantom}. For a scalar
field with a nearly flat potential, approximate relations between the equation of state parameter $w=p/\rho$ and
the energy density parameter $\Omega_\phi$ can be obtained \cite{Robert2008,Robert,Sourish,Crittenden:2007yy}. By using this generic relation,
more general dark energy models can be derived from known models with $\Omega_\phi$,
and the derived dark energy models can be approximated by the Chevallier-Polarski-Linder (CPL) parametrization \cite{cpl1,cpl2,Gong:2013bn}.
In this paper, we
focus on the thawing scalar field models with nearly flat potentials with a starting value of $w$ that is close to, but not
exactly equal to -1 at early times \cite{Caldwell:2005tm}.

Fundamental constants like the fine structure constant $\alpha=e^2/\hbar c$ play important roles in physics.
The variation of the fine structure constant $\alpha$ with time produce shifts in the molecular spectra \cite{Rodger}.
So observations of molecular spectra in high redshift objects can track the value of the fundamental constant in the early universe.
It is possible that the variation of the fine structure constant $\alpha$ is caused by the non-minimally coupling between dark energy
and the electromagnetic field strength \cite{Bekenstein}.
The measurements of the values of the fundamental constant throughout the history of the universe provide strong constraints on the property of dark energy \cite{Martins}.
A lot of efforts have been made to measure the variation of  the fundamental constants \cite{King2009,Wendt,Thompson,Malec,King2011,Murphy2008,Muller2011,Julian,Murphy:2003mi,Murphy2004,Chand2004,Webb}.
These observations can be used to test cosmological model \cite{zhangtj}.
On the other hand, the variations of the fundamental constants may caused by the
interaction of the vacuum with the matter \cite{Fritzsch:2012qc}. In particular,
it is possible that the vacuum energy is time-varying with $w_\Lambda=-1$ due
to the renormalization group running of the cosmological constant in the framework
of quantum field theory in curved space-time \cite{Shapiro:1999zt,Shapiro:2009dh,Basilakos:2009wi,Grande:2011xf}.

In this paper, we use the observational data on the variation of the fine structure constant $\Delta \alpha/\alpha$ to constrain the property of dark energy.
The paper is organized as follows. In section 2, we review the generic relationship between $w$ and $\Omega_\phi$ for both quintessence and phantom fields satisfying the  slow-roll conditions.
In section 3, we relate the variation of the fine structure constant $\Delta \alpha/\alpha$ with the thawing scalar field.
In section 4, we derive the thawing dark energy model and apply the observational data to constrain the model.
The conclusions are drawn in section 5.

\section{Slow-roll scalar fields}
Taking $w$ to be the ratio of pressure to energy density of the dark energy,
\begin{eqnarray}
\label{w}
w=p_{DE}/\rho_{DE},
\end{eqnarray}
models for which $w>-1$ are called quintessence \cite{wetterich88,peebles88,quintessence}
and models with $w<-1$ are called phantom \cite{phantom}.
Freezing models start with the equation of state parameter $w$ different from -1 at early times and approaching -1 at the present
time while thawing scalar field models start with $w$ close to -1 at early times and deviate from -1 at the present epoch \cite{Caldwell:2005tm}.
In this paper, we use a scalar field to model dark energy.

\subsection{Quintessence}
For the quintessence models, we assume that dark energy is provided by a minimally coupled scalar field $\phi$.
The pressure and energy density of the scalar field are given by
\begin{equation}
\label{p}
p_\phi=\frac{\dot{\phi}^2}{2}-V(\phi),
\end{equation}
and
\begin{equation}
\label{rho}
\rho_\phi=\frac{\dot{\phi}^2}{2}+V(\phi).
\end{equation}
The equation of motion of the scalr field is given by
\begin{equation}
\label{field}
\ddot{\phi} + 3 H \dot{\phi} + \frac{dV}{d\phi}=0,
\end{equation}
where the Hubble parameter $H$ is given by
\begin{equation}
\label{h}
H=\frac{\dot a}{a}=\kappa\,\sqrt{\rho/3}.
\end{equation}
Here $a$ is the scale factor, $\rho$ is the total energy density and $\kappa^2=8\pi G$.
We consider the spatially flat model only throughout the paper.

If the scalar field has a nearly flat potential $V(\phi)$ with initial value $\phi_0$,
i.e., at $\phi=\phi_0$, the scalar field satisfies the slow-roll conditions:
\begin{equation}
\label{slow1}
(\frac{1}{V}\frac{dV}{d\phi})^2\ll 1, \quad
\frac{1}{V}\frac{d^2V}{d{\phi}^2}\ll 1,
\end{equation}
then over the region in which the above conditions (\ref{slow1}) apply,  a generic relationship between $w$ and the energy
density $\Omega_\phi=\kappa^2\rho_\phi/3H^2$ for the quintessence was found \cite{Robert2008,Sourish},
\begin{equation}
\label{eq13}
1+w=\frac{\lambda_0 ^2}{3}\left[\frac{1}{\sqrt{\Omega_\phi}}-\left(\frac{1}{\Omega_\phi}-1\right)(\tanh^{-1}\sqrt{\Omega_\phi}+C)\right]^2,
\end{equation}
where $\lambda_0$ is the value of $\lambda=V^{-1}dV(\phi)/d\phi$ at the initial value
of the scalar field $\phi_0$ before it begins to roll down the potential,
and the integration constant $C$ is determined by the initial values $w_i$ and $\Omega_{\phi i}\ll 1$ \cite{Sourish}
\begin{equation}
\label{eq14}
C=\pm\frac{\sqrt{3(1+\omega_i)}\,\Omega_{\phi i}}{\lambda_0}.
\end{equation}
For simplicity, we neglect the early dark energy and take $C=0$ which corresponds to the thawing scalar field models.
As shown in Figures 2-4 with two explicit potentials $V(\phi)=\phi^2$ and $V(\phi)=\phi^{-2}$ in Ref. \refcite{Robert2008},
the generic relation (\ref{eq13}) between $w$ and $\Omega_\phi$ is a good approximation for arbitrary potentials
once the slow-roll conditions (\ref{slow1}) are satisfied.
In terms of the current values $w_0$ and $\Omega_{\phi 0}$, we get
\begin{eqnarray}
\label{lambda0}
\lambda_0=\frac{\sqrt{3(1+w_0)}}{\Omega_{\phi0}^{-1/2}-(\Omega_{\phi0}^{-1}-1)\tanh^{-1}(\sqrt{\Omega_{\phi0}})}.
\end{eqnarray}
Substituting Eq. (\ref{lambda0}) for $\lambda_0$ into Eq. (\ref{eq13}), we get
\begin{eqnarray}
\label{eq15}
1+w=(1+w_0)\left[\frac{1}{\sqrt{\Omega_{\phi0}}}-(\Omega_{\phi0}^{-1}-1)\tanh^{-1}
\sqrt{\Omega_{\phi0}}\right]^{-2}\nonumber\\
\times\left[\frac{1}{\sqrt{\Omega_\phi}}-\left(\frac{1}{\Omega_\phi}-1\right)\tanh^{-1}(\sqrt{\Omega_\phi})\right]^2.
\end{eqnarray}
Note that the above result does not depend on the specific form of the potential $V(\phi)$
and holds for general
potentials satisfying the slow-roll conditions (\ref{slow1}). Once a potential $V(\phi)$ is
given, we can solve the coupled cosmological equations to find the equation of state parameter $w$ for the scalar field.
However, with the result (\ref{eq15}),
we can get the function $w(z)$ from $\Omega_\phi$ without specifying the potential $V(\phi)$.
In other words, instead of specifying $V(\phi)$, we use $\Omega_\phi$ to obtain the equation of state $w(z)$
for the scalar field $\phi$, and it provides us with a  particular model for $w(a)$ by this way.

\subsection{Phantom}
For the phantom model, we consider a scalar field $\phi$ with negative kinetic term and potential $V(\phi)$,
the energy density and pressure of the phantom are given by
\begin{eqnarray}
\label{rho1}
\rho_\phi=-\frac{\dot{\phi}^2}{2}+V(\phi),
\end{eqnarray}
and
\begin{eqnarray}
\label{p1}
p_\phi=-\frac{\dot{\phi}^2}{2}-V(\phi),
\end{eqnarray}
so that the equation of state parameter is
\begin{eqnarray}
\label{w1}
w=\frac{\dot{\phi}^2+2V(\phi)}{\dot{\phi}^2-2V(\phi)}.
\end{eqnarray}
The evolution of $\phi$ is given by
\begin{eqnarray}
\label{field1}
\ddot{\phi} + 3 H \dot{\phi} - \frac{dV}{d\phi}=0.
\end{eqnarray}
Similarly, for the phantom field satisfying the slow-roll conditions, we get \cite{Robert}
\begin{eqnarray}
\label{w2}
1+w=-\frac{\lambda_0 ^2}{3}\left[\frac{1}{\sqrt{\Omega_\phi}}-\left(\frac{1}{\Omega_\phi}-1\right)(\tanh^{-1}\sqrt{\Omega_\phi}+C)\right]^2.
\end{eqnarray}
Again, we consider the case $C=0$ for simplicity. Expressing $\lambda_0$ in terms of $w_0$ and $\Omega_{\phi 0}$,
the result between $w$ and $\Omega_\phi$ for the phantom is the same as that for quintessence
given by equation (\ref{eq15}).

\section{The variation of fundamental constants}
For realistic dark energy models, the scalar field should couple to other matter components in the universe. Here we
consider the coupling between the scalar field and photon with the interaction \cite{Bekenstein},
\begin{equation}
\label{intereq1}
L_{\phi F}=-\frac{1}{4}B_F(\phi)F_{\mu\nu}F^{\mu\nu},
\end{equation}
where $B_F(\phi)=1-\zeta_\alpha\kappa(\phi-\phi_0)$  and the coupling constant $\zeta_\alpha$ is
constrained to be $|\zeta_\alpha|\sim10^{-4}-10^{-7}$ \cite{copeland04,Nunes}.
Due to the coupling, the fine structure $\alpha$ will evolve with the scalar field,
\begin{equation}
\label{alphaeq1}
\frac{\Delta \alpha}{\alpha}=\frac{\alpha-\alpha_0}{\alpha_0}=\zeta_\alpha\kappa(\phi-\phi_0).
\end{equation}
So
\begin{eqnarray}
\label{ophi}
\frac{\alpha'}{\alpha}=\zeta_\alpha\kappa \phi',
\end{eqnarray}
where $\alpha'=d\alpha/d \ln a={\dot\alpha}/H$. On the other hand,
\begin{eqnarray}
\label{phid}
|w+1|=\frac{\dot{\phi}^2}{\rho_\phi}=\frac{(\kappa\phi')^2}{3\Omega_\phi},
\end{eqnarray}
where $\phi'=d\phi/d \ln a={\dot\phi}/H$.
Substituting equation (\ref{ophi}) into equation (\ref{phid}), we get \cite{Rodger}
\begin{eqnarray}
\label{a1}
(\alpha'/\alpha)^2 =3\zeta_\alpha^2\, |w+1|\, \Omega_\phi.
\end{eqnarray}
Therefore,
\begin{eqnarray}
\label{dverk}
\left|\frac{\Delta\alpha}{\alpha}\right|&=&\int_1^a \sqrt{3\zeta_\alpha^2\Omega_\phi(x)|w(x)+1|}d\ln x \nonumber\\
&=&\int_0^z (1+z)^{-1}\,\sqrt{3\zeta_\alpha^2\,\Omega_\phi(z)|w(z)+1)|}\, dz.
\end{eqnarray}
For a model with known $w(z)$, we can solve the cosmological equations to get $\Omega_\phi$ and then calculate
the variation of the fine structure constant
$\Delta\alpha/\alpha$.

If $\Omega_\phi$ is given, substituting equation (\ref{eq15}) with the $\Omega_\phi$
into the above equation (\ref{dverk}), we get \cite{Rodger}
\begin{eqnarray}
\label{varalpha1}
\left|\frac{\Delta\alpha}{\alpha}\right|=\sqrt{3\zeta_\alpha^2|1+w_0|}\left|\frac{1}{\sqrt{\Omega_{\phi0}}}-(\Omega_{\phi0}^{-1}-1)\tanh^{-1}\sqrt{\Omega_{\phi0}}\right|^{-1}\nonumber\\
\times\int_0^z (1+z)^{-1} \left|1-\left(\Omega_\phi^{-1/2}-\sqrt{\Omega_\phi}\right)\tanh^{-1}(\sqrt{\Omega_\phi})\right|dz.
\end{eqnarray}

\section{Cosmological constraints}

\begin{figure}[pht]
\centerline{\includegraphics[width=84mm]{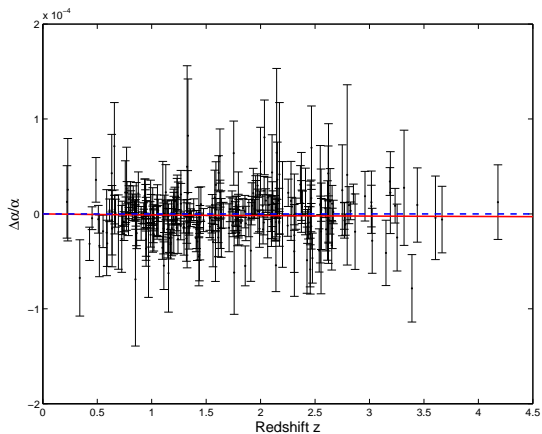}}
\caption{The 291 $\Delta\alpha/\alpha$ observational data \cite{Julian,Murphy:2003mi}.  The solid line is the evolution of $\Delta\alpha/\alpha$ as a
function of $z$ with equation (\ref{new}) for the best fitting value $\epsilon=4.2\times10^{-7}$.
The dashed line is the horizonal line indicating no variation of $\Delta\alpha/\alpha$.}
\label{fig1}
\end{figure}

Now we apply the observational data on $\Delta\alpha/\alpha$ to constrain the property of dark energy.
The data sample consists of 151 absorbers of quasar absorption-line spectra obtained using the Ultraviolet
and Visual Echelle Spectrograph on the Very Large Telescope in Chile \cite{Julian}, and 140 absorbers obtained with the Keck
telescope at the Keck Observatory in Hawaii \cite{Murphy:2003mi}.
We apply the $\chi^2$ statistics to the 291 data \cite{Julian,Murphy:2003mi},
\begin{eqnarray}
\label{chi}
\chi^2=\sum_i\left[\frac{(\Delta\alpha/\alpha)_{th,i}-(\Delta\alpha/\alpha)_{obs,i}}{\sigma_i}\right]^2,
\end{eqnarray}
where the subscripts $"th"$ and $"obs"$ stand for the theoretically predicted value and observed ones respectively.
The theoretical value of $\Delta\alpha/\alpha$ is calculated by equation (\ref{dverk}) or (\ref{varalpha1}).
Since $\Delta\alpha/\alpha$ is proportional to $\sqrt{\zeta_\alpha^2|1+w_0|}$, so the positive detection of the variation
of the fine structure constant means that dark energy is not a cosmological constant
if the variation of the fine structure is caused by dark energy with the interaction (\ref{intereq1}). However, if the variation of
$\alpha$ is not due to dark energy, then the above statement does not hold.
By neglecting the recent dark energy domination, an analytic expression for the behavior of $\alpha$ was proposed \cite{Vielzeuf},
\begin{eqnarray}
\label{new}
\frac{\Delta\alpha}{\alpha}=-4\epsilon\ln(1+z),
\end{eqnarray}
where $\epsilon$ gives the magnitude of the variation.
In order to get the best-fit result of the parameter $\epsilon$ in equation (\ref{new}),
we apply the $\chi^2$ statistics to the 291 $\Delta\alpha/\alpha$ observational data. The $1\sigma$ constraint on $\epsilon$ is
$\epsilon=(4.2\pm 2.3)\times10^{-7}$ with $\chi^2=296.48$. By using the best-fit value of $\epsilon$,
we plot the variation of $\Delta\alpha/\alpha$ in figure \ref{fig1}. If we choose $\epsilon=0$ which corresponds to the model with $\Delta\alpha/\alpha=0$,
we get $\chi^2=299.82$. The model with no variation of the fine structure is consistent with the data
at the $2\sigma$ level.
It is clear that $\alpha$ varies with redshift at the $1\sigma$ level. In the following, we apply the data to constrain
dark energy models.

\begin{figure}[pht]
\centerline{\includegraphics[width=84mm]{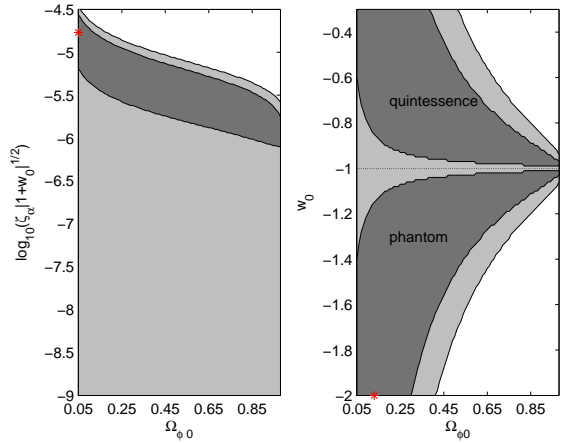}}
\caption{
The $1\sigma$ and $2\sigma$ contours on the dark energy model with $w(z)$ given by equation (\ref{wzeq1}).
Dark gray and light gray areas show the $1\sigma$ and $2\sigma$ contours, respectively. The contours between
$\Omega_{\phi 0}$ and $w_0$ in the right panel were obtained by fixing $|\zeta_\alpha|=1\times 10^{-5}$.}
\label{fig2}
\end{figure}

If $w$ is near $-1$, then $\Omega_\phi$ can be approximated with $\Lambda$CDM model \cite{Robert2008,Crittenden:2007yy},
\begin{eqnarray}
\label{phi}
\Omega_\phi=\frac{1}{1+(\Omega_{\phi 0}^{-1}-1)a^{-3}}.
\end{eqnarray}
Substituting equation (\ref{phi}) into equation (\ref{eq15}), we get \cite{Robert2008,Robert}
\begin{eqnarray}
\label{wzeq1}
w(a)=-1+(1+w_0)\left[\frac{1}{\sqrt{\Omega_{\phi0}}}-(\Omega_{\phi0}^{-1}-1)\tanh^{-1}
\sqrt{\Omega_{\phi0}}\right]^{-2}\nonumber\\
\times\left[\sqrt{1+(\Omega_{\phi 0}^{-1}-1)a^{-3}}-(\Omega_{\phi 0}^{-1}-1)a^{-3}\tanh^{-1}[1+(\Omega_{\phi 0}^{-1}-1)a^{-3}]^{-1/2}\right]^2.
\end{eqnarray}
In other words, we consider the dark energy model \cite{Robert2008,Robert} with
$w(a)$ given by equation (\ref{wzeq1}) and the model is different from $\Lambda$CDM model because $w(z)\neq -1$.
Taking Taylor expansion of $w(a)$ around $a=1$, the model can be approximated as the CPL parametrization at low redshift  with the degeneracy relationship
\begin{equation}
\label{waeq1}
w_a=6(1+w_0)\frac{\Omega_{\phi0}^{-1/2}-\sqrt{\Omega_{\phi 0}}-(\Omega_{\phi0}^{-1}-1)\tanh^{-1}(\sqrt{\Omega_{\phi0}})}
{\Omega_{\phi 0}^{-1/2}-(\Omega_{\phi0}^{-1}-1)\tanh^{-1}(\sqrt{\Omega_{\phi0}})}.
\end{equation}
If we take $\Omega_{\phi 0}=0.7$, then we get $w_a=-1.42(1+w_0)$
which is consistent with the numerical result $w_a\approx -1.5(1+w_0)$ obtained in Refs. \refcite{Robert2008,Robert}.
Note that we derived the analytical expression for $w_a$ in terms of both $\Omega_{\phi 0}$ and $w_0$.
With this explicit degeneracy relation, the CPL parametrization can be used to
give tighter constraints on $\Omega_{\phi 0}$ and $w_0$.
Following Ref. \refcite{Rodger}, we consider the model (\ref{waeq1}) and substitute equation (\ref{phi}) into equation (\ref{varalpha1})
to get the change of the fine structure constant. By using
the observational data, we get constraints on $\log_{10}( \sqrt{\zeta_\alpha^2 |1+w_0|})$ and $\Omega_{\phi0}$
and the results are shown in Figure \ref{fig2}.
The best fitting results are $\log_{10}(\sqrt{\zeta_\alpha^2 |1+w_0|})=-4.77$ and $\Omega_{\phi0}=0.05$ with $\chi^2=293.97$.
At the $1\sigma$ level, $\Omega_{\phi 0}$ reaches the physical boundaries 0 and 1, and $\log_{10}(\sqrt{\zeta_\alpha^2 |1+w_0|})=-4.77^{+0.21}_{-0.43}$.
If we take the observational value $\Omega_{\phi0}=0.72$ \cite{gong13}, then the $1\sigma$ constraint is
$\log_{10}(\sqrt{\zeta_\alpha^2|1+w_0|})=-5.5^{+0.1}_{-0.3}$ with $\chi^2=294.32$.
It is clear that the data is not sensitive to $\Omega_{\phi 0}$ and bigger value of $\zeta_\alpha^2|1+w_0|$ is needed
to compensate smaller $\Omega_{\phi 0}$.
Taking the observational value $|1+w_0|=0.02$ \cite{gong13},
the best fitting $\zeta_\alpha$ is roughly equal to $1\times10^{-5}$.
By using the best fitting value $|\zeta_\alpha|=1\times10^{-5}$, and the priors
$-2.0\leqslant w_0 \leqslant -0.3$ and $0.05\leqslant \Omega_{\phi0} \leqslant 0.99$,
we get constraints on $w_0$ and $\Omega_{\phi0}$,
and the best fitting values are $\Omega_{\phi0}=0.13$ and $w_0=-2.0$ with $\chi^2=293.99$.
Figure \ref{fig2} shows the $1\sigma$ and $2\sigma$ contours on
$\Omega_{\phi0}$ and $w_0$.

\section{Conclusions}
Starting with a generic relationship (\ref{eq15}) between $w$ and $\Omega_\phi$ for thawing scalar field models,
we discussed the dark energy model with varying $w(z)$ by using $\Omega_\phi$ from $\Lambda$CDM.
Since at the $1\sigma$ level,
the fine structure constant is found to vary with redshift according to $-4\epsilon \ln (1+z)$
as shown in Figure \ref{fig1}, so we apply the data of $\Delta \alpha/\alpha$ to constrain
the property of thawing scalar field models
with the assumption that the variation of the fine structure constant is caused by dark energy.

The variation of the fine structure constant $\Delta\alpha/\alpha$ is proportional to the coupling constant $\zeta_\alpha$
by the factor $\sqrt{\zeta_\alpha^2|1+w|}$,
so there are three parameters $\zeta_\alpha$, $w_0$ and $\Omega_{\phi 0}$ to be fitted.
However, the data only constrains $\Omega_{\phi 0}$ and $\sqrt{\zeta_\alpha^2|1+w_0|}$. If we can determine
the values of $w_0$ and $\Omega_{\phi 0}$ from other astronomical observations, then the 291 data of $\Delta\alpha/\alpha$
can be used to constrain the coupling constant $\zeta_\alpha$. If we take the observational
values $\Omega_{\phi 0}=0.72$ and $|1+w_0|=0.02$, we get the best fitting value $|\zeta_\alpha|=1\times 10^{-5}$.
Because $\Delta\alpha/\alpha$ is proportional to $\sqrt{\zeta_\alpha^2|1+w_0|}$, so $\sqrt{\zeta_\alpha^2|1+w_0|}$
determines the magnitude of $\Delta\alpha/\alpha$ and is independent of the variation of $\Delta\alpha/\alpha$ with redshift,
we need to find other ways to break the the degeneracy between $\zeta_\alpha$ and $w_0$. A bigger value of $|1+w_0|$
is needed if smaller value of $\zeta_\alpha$ is chosen, that is the reason why the best-fit value
of $w_0$ approaches the cut-off value $w_0=-2$ and the best-fit value of $\Omega_{\phi 0}$ is small.
However, the value of $\chi^2$ does not change much when we choose the observational values $|1+w_0|=0.02$ and $\Omega_{\phi 0}=0.72$
instead of the  best-fit values obtained here.
Due to the big uncertainties of the measurements, the observational constraints are not tight.
The 291 data of $\Delta\alpha/\alpha$ excluded some areas around $w_0=-1$ at the $1\sigma$ level,
$w_0$ is unbounded from below for phantom fields and unbounded from above for quintessence fields.
Because $w_0=-1$ gives $\Delta\alpha/\alpha=0$, so positive detection of the variation of the fine
structure at the $1\sigma$ level means $w_0\neq -1$ at the $1\sigma$ level if the variation
of the fine structure is caused by dark energy with the interaction (\ref{intereq1}). That is why we get an excluded area around $w_0=-1$.
However, $\Lambda$CDM model is consistent with the data at $2\sigma$ level.

\section*{Acknowledgement}
The authors would like to thank Tong-Jie Zhang for helpful discussions. 
This work was partially supported by
the National Basic Science Program (Project 973) of China under
grant No. 2010CB833004, the NNSF of China under grant Nos. 10935013 and 11175270,
the Program for New Century Excellent Talents in University
and the Fundamental Research Funds for the Central Universities.


\end{document}